\documentclass[twocolumn,amsmath,amssymb,showpacs]{revtex4}
\usepackage{graphicx}

\begin{document}


\title{Generating the Kochen-Specker Sets with 36, 38 and 40 Rays in Three-Qubit System: Three Simple Algorithms}


\author{S.P.Toh\footnote{SingPoh.Toh@nottingham.edu.my; singpoh@gmail.com}}

\affiliation {Faculty of Engineering, The University of Nottingham Malaysia Campus, 
Jalan Broga, 43500 Semenyih, Selangor Darul Ehsan, Malaysia.}


\date{\today}

\begin{abstract}
\emph{We put forward three simple algorithms to generate the
Kochen-Specker sets used for the parity proof of the Kochen-Specker theorem
in three-qubit system. These algorithms enable us to generate 320,
640 and 64 Kochen-Specker sets with 36, 38 and 40 rays,
respectively. No any computer calculation is required, every step in the algorithms
is determined by the method of picking the rays that repeat 4 times in the KS sets.}
\end{abstract}


\maketitle


Besides nonlocality, contextuality is also a fascinating property of
quantum mechanics. In the contextual quantum world, the outcome of a
measurement depends on which compatible observables are
measured together. A landmark statement of contxtuality is the
Kochen-Specker (KS) theorem which asserts that quantum mechanics can
be completed only by contextual hidden variable model. Specifically,
KS theorem states that in a Hilbert space of dimension $d > 2$, it
is impossible to associate definite numerical values, 1 or 0, with
every rays (vectors) in a finite set, in such a way that, (1) no
two orthogonal rays are both assigned the value 1, and (2) none of the
complete basis is assigned the value 0 to all its rays.

The first explicit proof of the KS theorem in 1967 used 117 rays
\cite{R1}. The number of rays is greatly reduced to 33 and 24 for
three and four dimensions, respectively \cite{R2}. Conceiving a proof with
the lowest number of rays is a topic under intensive research. The current records
for three-, four-, five-, six-, seven- and eight-dimensional systems are
31 rays \cite{R3}, 18 rays \cite{R4}, 29 rays \cite{R5}, 31 rays \cite{R5},
34 rays \cite{R5} and 36 rays \cite{R6}, respectively.

There are also magnificent progresses made in the experiments of
testing KS theorem. In 1999, Meyer and Kent \cite{R7,R8} started a
discussion on the possibility of testing KS theorem experimentally,
for the proof proposed theoretically relied on infinite precision of
measurement. By taking into account the unavoidable imprecisions that
exist in actual experiments, Cabello \cite{R9} proposed a KS
inequality that is satisfied by a noncontextual hidden-variables theory
but is violated by quantum mechanics. Subsequently, the violation of
the KS inequality has been shown in the experiments using trapped
ions \cite{R10}, single photons \cite{R11} and nuclear magnetic
resonance system \cite{R12}. The inequality introduced in \cite{R9}
is universal, i.e., any KS set in dimension $d\geq3$ can be converted
into a noncontextuality inequality which is state-independently
violated \cite{R13}.

Efforts of physicists no longer restricted on finding various
proofs of KS theorem, the connection between the KS and Bell's theorems
becomes one of the interesting research topics. Contextuality is a more general
concept than nonlocality and it has been shown that
every KS set can derive a maximally violated Bell inequality \cite{R14}.
The state-independent noncontextuality inequalities provides a way to
test the nonlocality of quantum system \cite{R15}. Besides, the investigation
on KS theorem also extended to its application on quantum cryptography
\cite{R16,R17}.

Recently, there are some progresses on the proof of KS theorem in higher dimensions \cite{R18,R19}.
For eight-dimensional three-qubit system, \cite{R18,R20}
reported the numbers of parity proofs using 36, 38 and 40 rank-1 projectors, i.e.,
320, 640 and 64, respectively. We will propose three simple
algorithms to generate these KS sets. In some cases \cite{R4,R21,R22} , the generation and investigation
of huge number of KS sets rely on computer program, our algorithms have advantages of generating
1024 KS sets quick without involving any computer calculation and can be followed easily even by non-physicists.
The similar beauty of simplicity is presented in \cite{R23} for four-dimensional KS sets.

This Letter is organized as follows. We will firstly introduce 40 sets of 4-rays.
We discuss two properties of these sets of rays and their roles in generating bases required
to form a KS set. The algorithms of generating eight
dimensional KS sets using 36, 38 and 40 rays are then explicated with the aid of examples. The
numbers of KS sets that can be generated by these algorithms are
also explained based on the steps proposed. Summary given as last portion of this Letter.

A Kochen-Specker (KS) set is a set of rays and bases used to prove the
KS theorem. A KS set can be used to perform parity proof if the
number of bases is odd and each ray occurs in an even number of times
among the bases \cite{R20}. Table 3 in \cite{R20} gives the types and
numbers of KS sets in three-qubit system that can be used for
performing parity proof. Adopting the Algorithms I, II and III we
propose, the three different types of the KS sets
can easily be generated.

Table \ref{T1} shows all the 25 bases used in the proof of Kernaghan
and Peres \cite{R6} for a system of three qubits. The bases are
formed by the 40 rays denoted as $R_i$, with \mbox{$i=$1, 2, 3, \textellipsis, 40},
which are explicitly listed in Table 1 of \cite{R6}. Since we will heavily use the 
Table 2 of \cite{R20}, we reproduce it as Table \ref{T1} here.  The first 5 bases are pure bases ($PB_i$,
$1 \leq i \leq 5$) and the rest are hybrid bases ($HB_i$, $6 \leq i \leq 25$). Hybrid bases
are formed by an equal mixture of rays from a pair of pure bases
\cite{R20}. It can be easily seen from Table \ref{T1} that the eight
dimensional 40 rays occur once in $PB$ and 4 times in $HB$.

\begin{table}[htb]
\caption{The 25 bases formed by the 40 eight dimensional rays.\label{T1}}
\begin{center}
\begin{tabular}{|c|cccccccc|}
  \hline
  \multicolumn{1}{|c|}{Index}& \multicolumn{8}{|c|}{Rays in Basis} \\
  \hline
  1 & 1 & 2 & 3 & 4 & 5 & 6 & 7 & 8 \\
  2 & 9 & 10 & 11 & 12 & 13 & 14 & 15 & 16 \\
  3 & 17 & 18 & 19 & 20 & 21 & 22 & 23 & 24 \\
  4 & 25 & 26 & 27 & 28 & 29 & 30 & 31 & 32 \\
  5 & 33 & 34 & 35 & 36 & 37 & 38 & 39 & 40 \\ \hline
  6 & 1 & 2 & 3 & 4 & 13 & 14 & 15 & 16 \\
  7 & 1 & 2 & 5 & 6 & 21 & 22 & 23 & 24 \\
  8 & 1 & 3 & 5 & 7 & 29 & 30 & 31 & 32 \\
  9 & 1 & 4 & 6 & 7 & 37 & 38 & 39 & 40 \\
  10 & 2 & 3 & 5 & 8 & 33 & 34 & 35 & 36 \\
  11 & 2 & 4 & 6 & 8 & 25 & 26 & 27 & 28 \\
  12 & 3 & 4 & 7 & 8 & 17 & 18 & 19 & 20 \\
  13 & 5 & 6 & 7 & 8 & 9 & 10 & 11 & 12 \\
  14 & 9 & 10 & 13 & 14 & 19 & 20 & 23 & 24 \\
  15 & 9 & 11 & 13 & 15 & 27 & 28 & 31 & 32 \\
  16 & 9 & 12 & 14 & 15 & 34 & 36 & 38 & 39 \\
  17 & 10 & 11 & 13 & 16 & 33 & 35 & 37 & 40 \\
  18 & 10 & 12 & 14 & 16 & 25 & 26 & 29 & 30 \\
  19 & 11 & 12 & 15 & 16 & 17 & 18 & 21 & 22 \\
  20 & 17 & 19 & 21 & 23 & 26 & 28 & 30 & 32 \\
  21 & 17 & 20 & 22 & 23 & 35 & 36 & 37 & 39 \\
  22 & 18 & 19 & 21 & 24 & 33 & 34 & 38 & 40 \\
  23 & 18 & 20 & 22 & 24 & 25 & 27 & 29 & 31 \\
  24 & 25 & 28 & 30 & 31 & 33 & 36 & 37 & 38 \\
  25 & 26 & 27 & 29 & 32 & 34 & 35 & 39 & 40 \\
  \hline
\end{tabular}
\end{center}
\end{table}

The three simple algorithms we propose in this Letter generate
the bases to form the KS sets from Table \ref{T1}. By generating the bases we mean picking
appropriate bases from Table \ref{T1} that satisfy some prescribed
conditions in the algorithms. We illustrate the algorithms by examples. Readers could check
the expected features accordingly.

The algorithms explicated in this Letters require the use of 4 rays,
$\Gamma^{ij} = \{\alpha, \beta, \gamma, \delta\}$, from the $PB$ that
repeat its 3-element subsets $\Gamma^{ij}_1=\{\alpha, \beta,
\gamma \}$, $\Gamma^{ij}_2=\{\alpha, \beta, \delta \}$,
$\Gamma^{ij}_3=\{\alpha, \gamma, \delta \}$ and
$\Gamma^{ij}_4=\{\beta, \gamma, \delta \}$ in $HB$. We name such
property possessed by a specific set of 4 rays as Property I.
There are 8 sets of 4-rays having Property I in each $PB$ and
all the 40 $\Gamma^{ij}$ with $1 \leq i \leq 5$ (subscript of $PB$) and
$1 \leq j \leq 8$ (subscript of the 4-rays) are listed in Table
\ref{T2}. Note that the elements of $\Gamma^{ij}$ are listed in the
ascending order. The complement of $\Gamma^{ij}$ with respect to the
$PB_i$ is $^{\neg} \Gamma^{ij} = PB_i-\Gamma^{ij}$, where the rays are
again listed in ascending order. In our algorithms, steps proposed based
on the sequence, $k$ of the rays repeating 4 times, denoted as $\Sigma_k$,
be chosen. All the ray in $\Gamma^{ij}$ chosen are $\Sigma_k$, but the
converse is not true.

Due to Property I, each choice of $\Gamma^{ij}$ will
generate 1 $PB$ and 4 $HB$s, we call this result of
choosing $\Gamma^{ij}$ as Property II. Both Properties I and II
guarantee the 4 rays in $\Gamma^{ij}$ occur 4 times, whereas
the rays of $^{\neg} \Gamma^{ij}$ in the corresponding $PB_i$ repeat for a
second time in the $HB$s generated. The rays that occur once in the
4 $HB$s are labeled as $\Lambda_i$, with the subscript denotes which
$HB$ it is obtained from. Each of the $\Lambda_i$ has 4
elements and they play important roles in determining $\Sigma_k$ in
the subsequent steps.

It is worthy to give an explicit example. Picking $\Gamma^{34}=\{R17, R21,
R22, R23\}$ from Table \ref{T2} generates 1 $PB$ and 4 $PB$s as shown in Table \ref{T3}. The
boldface rays in $HB_{19}$, $HB_{20}$ and $HB_{21}$ and $HB_{7}$
constitute $\Gamma^{34}_1$, $\Gamma^{34}_2$, $\Gamma^{34}_3$ and
$\Gamma^{34}_4$, respectively. The rays $^{\neg} \Gamma^{34} = \{R18,
R19, R20, R24\}$ from $PB_3$ are italicized, and they occur for the second time in
$HB_{19}$, $HB_{20}$ and $HB_{21}$ and $HB_{7}$, respectively. In Table \ref{T3},
it is obvious that $\Lambda_{19}=\{R11, R12, R15, R16\}$, $\Lambda_{20}=\{R26, R28, R30, R32\}$,
$\Lambda_{21}=\{R35, R36, R37, R39\}$ and $\Lambda_7=\{R1, R2, R5, R6\}$.

In the cases where more than one $\Gamma^{ij}$ are required, as what would happen
in both Algorithm II and Algorithm III, one particular $PB$ can offer one and only one
$\Gamma^{ij}$ from Table \ref{T2}. Therefore, for example, while two chosen $\Gamma^{ij}$
will definitely generate 2 different $PB$s, the total number of $HB$ generated would be less than 8,
if some of them are overlapping.  We will pick only once for the
overlapping $HB$. The number of $HB$ generated determines the number of set for
$\Lambda_i$.

Finding $\Gamma^{ij}$ and $\Lambda_i$ is essential in all the three algorithms. In Algorithm II
and Algorithm III, we also need to find the subsets of $\Lambda_i$, and the intersections of
 $\Lambda_i$ (denoted as $\Delta_i$ and $\Xi^i_j$, respectively in the following). With the aid of these 4
types of the sets of rays, the three algorithms enable us to generate completely the KS sets
with 36, 38 and 40 rays for three-qubit system.

\begin{table*}[htb]
\caption{Sets of four rays possessing Property I and Property II.
These sets are all taken from pure bases.} \label{T2}
\begin{center}
\begin{tabular}{|c|c|c|c|c|c|}
  \hline
  $j$ & $PB_1$ & $PB_2$ & $PB_3$ & $PB_4$ & $PB_5$ \\ \hline
  1 & 1\hspace{2mm}2\hspace{2mm}3\hspace{2mm}5  & \hspace{2mm}9\hspace{2mm}10\hspace{2mm}11\hspace{2mm}13  & 17\hspace{2mm}18\hspace{2mm}19\hspace{2mm}21 & 25\hspace{2mm}26\hspace{2mm}27\hspace{2mm}29 & 33\hspace{2mm}34\hspace{2mm}35\hspace{2mm}40 \\
  2 & 1\hspace{2mm}2\hspace{2mm}4\hspace{2mm}6  & \hspace{2mm}9\hspace{2mm}10\hspace{2mm}12\hspace{2mm}14  & 17\hspace{2mm}18\hspace{2mm}20\hspace{2mm}22 & 25\hspace{2mm}26\hspace{2mm}28\hspace{2mm}30 & 33\hspace{2mm}34\hspace{2mm}36\hspace{2mm}38 \\
  3 & 1\hspace{2mm}3\hspace{2mm}4\hspace{2mm}7  & \hspace{2mm}9\hspace{2mm}11\hspace{2mm}12\hspace{2mm}15  & 17\hspace{2mm}19\hspace{2mm}20\hspace{2mm}23 & 25\hspace{2mm}27\hspace{2mm}28\hspace{2mm}31 & 33\hspace{2mm}35\hspace{2mm}36\hspace{2mm}37 \\
  4 & 1\hspace{2mm}5\hspace{2mm}6\hspace{2mm}7  & \hspace{2mm}9\hspace{2mm}13\hspace{2mm}14\hspace{2mm}15  & 17\hspace{2mm}21\hspace{2mm}22\hspace{2mm}23 & 25\hspace{2mm}29\hspace{2mm}30\hspace{2mm}31 & 33\hspace{2mm}37\hspace{2mm}38\hspace{2mm}40 \\
  5 & 2\hspace{2mm}3\hspace{2mm}4\hspace{2mm}8  & 10\hspace{2mm}11\hspace{2mm}12\hspace{2mm}16 & 18\hspace{2mm}19\hspace{2mm}20\hspace{2mm}24 & 26\hspace{2mm}27\hspace{2mm}28\hspace{2mm}32 & 34\hspace{2mm}35\hspace{2mm}36\hspace{2mm}39 \\
  6 & 2\hspace{2mm}5\hspace{2mm}6\hspace{2mm}8  & 10\hspace{2mm}13\hspace{2mm}14\hspace{2mm}16 & 18\hspace{2mm}21\hspace{2mm}22\hspace{2mm}24 & 26\hspace{2mm}29\hspace{2mm}30\hspace{2mm}32 & 34\hspace{2mm}38\hspace{2mm}39\hspace{2mm}40 \\
  7 & 3\hspace{2mm}5\hspace{2mm}7\hspace{2mm}8  & 11\hspace{2mm}13\hspace{2mm}15\hspace{2mm}16 & 19\hspace{2mm}21\hspace{2mm}23\hspace{2mm}24 & 27\hspace{2mm}29\hspace{2mm}31\hspace{2mm}32 & 35\hspace{2mm}37\hspace{2mm}39\hspace{2mm}40 \\
  8 & 4\hspace{2mm}6\hspace{2mm}7\hspace{2mm}8  & 12\hspace{2mm}14\hspace{2mm}15\hspace{2mm}16 & 20\hspace{2mm}22\hspace{2mm}23\hspace{2mm}24 & 28\hspace{2mm}30\hspace{2mm}31\hspace{2mm}32 & 36\hspace{2mm}37\hspace{2mm}38\hspace{2mm}39 \\
  \hline
\end{tabular}
\end{center}
\end{table*}

\begin{table}[htb]
\caption{Due to Properties I and II, the chosen $\Gamma^{34}$ generates $PB_3$,
$HB_{19}$, $HB_{20}$, $HB_{21}$ and $HB_7$.  From the 5 bases, it is easy to
obtain sets for rays that occur once ($\Lambda_{19}$, $\Lambda_{20}$, $\Lambda_{21}$
and $\Lambda_7$), twice ($^{\neg} \Gamma^{34}$) and four times ($\Gamma^{34})$.} \label{T3}
\begin{center}
\begin{tabular}{|c|cccccccc|}
  \hline
  \multicolumn{1}{|c|}{Index}& \multicolumn{8}{|c|}{Rays in Basis} \\
  \hline
  3 & $\bf{17}$ & $\it{18}$ & $\it{19}$ & $\it{20}$ & $\bf{21}$ & $\bf{22}$ & $\bf{23}$ & $\it{24}$ \\
  7 & 1 & 2 & 5 & 6 & $\bf{21}$ & $\bf{22}$ & $\bf{23}$ & $\it{24}$ \\
  19 & 11 & 12 & 15 & 16 & $\bf{17}$ & $\it{18}$ & $\bf{21}$ & $\bf{22}$ \\
  20 & $\bf{17}$ & $\it{19}$ & $\bf{21}$ & $\bf{23}$ & 26 & 28 & 30 & 32 \\
  21 & $\bf{17}$ & $\it{20}$ & $\bf{22}$ & $\bf{23}$ & 35 & 36 & 37 & 39 \\
  \hline
\end{tabular}
\end{center}
\end{table}

Lets begin with Algorithm I. The following steps generate the KS sets of the type $28_28_4\textendash11_8$.
The symbol $28_28_4\textendash11_8$ means that 28 of the rays
occur twice and 8 of the rays occur 4 times in 11 bases \cite{R20}. The
subscript 8 means that there are 8 rays in each basis. The 11 bases
consist of 1 $PB$ and 10 $HB$s. The existence of 1 $PB$ means
that only 1 $\Gamma^{ij}$ is required. Algorithm I shows us how to
pick the 8 rays that occur 4 times, denoted $\Sigma_k$ ($k=1, 2, \textellipsis, 8$),
in such a way that the other 28 rays in the 11 bases occur twice.

{\bf Step 1}: Pick the first $\Gamma^{ij}$ from Table \ref{T2}.

Choosing $\Gamma^{ij}$ means choosing $\Sigma_1 = \alpha$,
$\Sigma_2=\beta$, $\Sigma_3=\gamma$ and $\Sigma_4=\delta$ (Property I). As
mentioned previously, $\Gamma^{ij}$ incurs 1 $PB$ and 4 $HB$s
(Property II). In addition, there are 4 rays of $^{\neg}
\Gamma^{ij}$ and 4 sets of $\Lambda^n_i$, with $n=1$ here. We
add the superscript $n$ to indicate that the corresponding $\Lambda_i$
are produced after the execution of Step $n$.

As an example, if we choose $\Gamma^{11}$, we have
$\Sigma_1=\alpha=R1$, $\Sigma_2=\beta=R2$, $\Sigma_3=\gamma=R3$ and
$\Sigma_4=\delta=R5$. To satisfy Property II, we need to choose $HB_6$,
$HB_{7}$, $HB_{8}$ and $HB_{10}$ as well as $PB_1$. Thus we obtain
5 bases after executing Step 1. The set of rays that repeat
twice in these 5 bases is $^{\neg} \Gamma^{11}=\{R4, R6, R7, R8\}$.
We remind ourselves that the rays in $\Lambda^1_i$ occur only once in the 5 bases
generated. The choice of $\Sigma_5$ will come from
$\Lambda^1_6=\{R13, \textellipsis, R16\}$, $\Lambda^1_7=\{R21,
\textellipsis, R24\}$, $\Lambda^1_8=\{R29, \textellipsis, R32\}$ or
$\Lambda^1_{10}=\{R33, \textellipsis, R36\}$.

{\bf Step 2}: Pick a ray in one of the $\Lambda^1_i$ as $\Sigma_5$.
Denote the corresponding $\Lambda^1_i$ as $\Lambda^1_a$.

The ray $\Sigma_5$ will occur 4 times in $HB$. Since one of them already
existed in $\Lambda^1_a$, there are only 3 new $HB$s generated
after the execution of Step 2. Note that it remains 3 $\Lambda^1_i (i \neq a)$ and
since they are 3 newly generated $HB$s, they are also 3 $\Lambda^2_i$. Three pairs
of rays, i.e. one pair from each of $\Lambda^1_i (i \neq a)$, must repeat themselves
in $\Lambda^2_i$. We denote them as $\Delta_i$, with the $i$ takes on
the same subscript as in $\Lambda^1_i$.

In our example, we choose $R13$ from $\Lambda^1_6$ as $\Sigma_5$. The other
$HB$s that contain $R13$ are $HB_{14}$, $HB_{15}$ and $HB_{17}$. Note that $\Delta_7 \subset
\Lambda^1_7 = \{R23, R24\}$, $\Delta_8 \subset \Lambda^1_8 = \{R31,
R32\}$ and $\Delta_{10} \subset \Lambda^1_{10} = \{R33, R35\}$ occur
also in $HB_{14}$, $HB_{15}$ and $HB_{17}$, respectively.

We will pick $\Sigma_6$, $\Sigma_7$ and $\Sigma_8$ from the 3 $\Delta_i$ in the
subsequent 3 steps.

{\bf Step 3}: There are 3 $\Delta_i$ produced by Step 2 and each of them contains 2 rays.
Pick one of the rays from any one of $\Delta_i$ as $\Sigma_6$.

As $\Sigma_6$ is chosen from $\Delta_i$, that means it already existed twice.
Thus Step 3 generates only 2 $HB$s. In our example, we execute
Step 3 by taking the $R23$ from $\Delta_7$ as $\Sigma_6$. The second
$R23$ existed in $HB_{14}$. The third and fourth $R23$ exist in
$HB_{20}$ and $HB_{21}$, thus $HB_{20}$ and $HB_{21}$ are newly generated bases.

{\bf Step 4}: Consider either one of the remaining 2 $\Delta_i$, only 1 ray of it
will generates a new $HB$. Take it as $\Sigma_7$.

It remains $\Delta_8$ and $\Delta_{10}$ in our example. Lets consider $\Delta_8$.
Thus, $R32$ is $\Sigma_7$ for it generates only $HB_{25}$, the first 3 rays of
$R32$ existed in $HB_8$, $HB_{15}$ and $HB_{20}$. The ray $R31$ in
$\Delta_8$ generates no new $HB$, so it can not be chosen as $\Sigma_7$.

{\bf Step 5}: Pick those ray from the last $\Delta_i$ that occurred 4 times
and without generating any new $HB$ as $\Sigma_8$.

It can be easily seen that $R35$ from $\Delta_{10}$ should be taken as $\Sigma_8$
in our example because it existed 4 times in the previously generated $HB$.

In our example, the 11 bases generated are $PB_1$, $HB_6$, $HB_7$,
$HB_8$, $HB_{10}$, $HB_{14}$, $HB_{15}$, $HB_{17}$, $HB_{20}$,
$HB_{21}$ and $HB_{25}$ (following the order in which they are generated).
The 8 rays that occur 4 times are $R1$, $R2$, $R3$, $R5$, $R13$, $R23$,
$R32$ and $R35$ (following the order in which they are chosen). The
remaining 28 rays occur twice in the generated 11 bases.

For Algorithm I, there are 40 ways of choosing the first 4
$\Sigma_i$ from $\Gamma^{ij}$ in Step 1, 4 ways of choosing
$\Sigma_5$ in Step 2 and 2 ways of choosing $\Sigma_6$ in Step 3,
therefore the total number of KS sets in the form of
$28_28_4\textendash11_8$, $N_{I}$, that can be generated by
Algorithm I is $40\times 4 \times 2 = 320$. Note that because $\Sigma_7$ and
$\Sigma_8$ in Step 4 and Step 5 are restricted by the $\Sigma_6$
chosen in Step 3, they don't contribute to $N_{I}$.

We now explicates the steps taken in Algorithm II to generate KS sets of the type
$24_214_4\textendash13_8$. The 13 bases consist of 3 $PB$s and 10
$HB$s. The existence of 3 $PB$s means that 3 $\Gamma^{ij}$ are required.
Algorithm II shows us how to pick the 14 rays that occurs
4 times, $\Sigma_k$ ($k=1, 2, \textellipsis, 14$), in such a way that the
remaining 24 rays occur twice in the KS sets generated.

{\bf Step 1}: Pick the first $\Gamma^{ij}$ from Table \ref{T2}.

This step is the same as Step 1 in Algorithm I. We take the same example
as for Step 1 of Algorithm I.

{\bf Step 2}: Pick the second $\Gamma^{ij}$ based on one of the
$\Lambda^1_i$.

By adding one appropriate ray, it can be easily seen from Table \ref{T2}
that any 3 rays from $\Lambda^1_i$ offer us options to choose
the second $\Gamma^{ij}$. Denotes the chosen $\Lambda^1_i$ as
$\Lambda^1_{a}$. Label the 4 rays in the second $\Gamma^{ij}$ as
$\Sigma_5, \textellipsis, \Sigma_8$. Three $\Lambda^2_i$ are obtained as Step 2
generates only 3 $HBs$. Define non-empty sets
$\Xi^i_j=\Lambda^1_i(i \neq a) \cap \Lambda^2_j$ and its complement
$^{\neg} \Xi^i_j=(\Lambda^1_i(i \neq a) \cup \Lambda^2_j)- \Xi^i_j$. Note that
$\Xi^i_j$ and $^{\neg} \Xi^i_j$ have 2 and 4 elements, respectively.

In our example, by adding $R9$, $R10$, $R11$ or $R12$,
$\Lambda^1_6=\{R_{13},\textellipsis, R_{16}\}$ allows us to choose
$\Sigma_5$, \textellipsis, $\Sigma_8$ either from $\Gamma^{24}$, $\Gamma^{26}$,
$\Gamma^{27}$, or $\Gamma^{28}$, respectively (see Table \ref{T2}). We now take
$\Gamma^{24}$ to generate 1 $PB$ and 3 $HB$s, i.e., $PB_2$,
$HB_{14}$, $HB_{15}$ and $HB_{16}$. Note that since both $\Gamma^{11}$ in Step 1 and
$\Gamma^{24}$ in Step 2 generate $HB_6$, only 3 (not 4 as in Step 1) $HB$s are
generated from the execution of Step 2 (see aforementioned Property II).
Excluding $\Lambda^1_{a}$, with $a=6$, we now have 3 $\Lambda^1_{i}$, i.e.,
$\Lambda^1_7=\{R21, \textellipsis, R24\}$, $\Lambda^1_8=\{R29, \textellipsis, R32\}$, and
$\Lambda^1_{10}=\{R33, \textellipsis, R36\}$. On the other hand, we have $\Lambda^2_{14}=\{R19, R20, R23, R24\}$,
$\Lambda^2_{15}=\{R27, R28, R31, R32\}$ and $\Lambda^2_{16}=\{R34, R36, R38, R39\}$.
By taking intersection, we thus obtain $\Xi^7_{14}=\{R23, R24\}$, $\Xi^8_{15}=\{R31, R32\}$ and $\Xi^{10}_{16}=\{R34, R36\}$.
On the other hand, $^{\neg} \Xi^7_{14}=\{R19, \textellipsis, R22\}$, $^{\neg}
\Xi^8_{15}=\{R27, \textellipsis, R30\}$ and $^{\neg}\Xi^{10}_{16}=\{R33, R35, R38, R39\}$.

{\bf Step 3}: Referring to Table \ref{T2}, it is obvious that the
rays from each pair of $\Xi^i_j$ and $^{\neg} \Xi^i_j$ offer two
options of taking the third $\Gamma^{ij}$. Pick one of it to form
$\Sigma_9, \textellipsis, \Sigma_{12}$.

Execution of Step 3 generates 1 $PB$ and 2 $HB$s, and 2
$\Lambda^3_i$ are obtained accordingly. Note that the number of sets for both
$\Lambda^1_i$ and $\Lambda^2_i$ reduced from 3 to 2 after the
execution of Step 3 and we denote them as $\Lambda^{1'}_i$ and
$\Lambda^{2'}_i$, respectively.

In our example, we take $\Xi^7_{14}$ and the rays $R19$ and $R21$
from $^{\neg} \Xi^7_{14}$ to form $\Gamma^{37}$, thus $\Sigma_9 = R19$,
$\Sigma_{10}=R21$, $\Sigma_{11}=R23$ and $\Sigma_{12}=R24$. Due to
this choice, only $PB_3$, $HB_{20}$ and $HB_{22}$ are generated, as
$HB_7$ and $HB_{14}$ that contain $\{R21, R23, R24\}$ and $\{R19,
R23, R24\}$ were existed. As a result of choosing $\Xi^7_{14}$, $\Lambda^1_7$
and $\Lambda^2_{14}$ have to be excluded in the subsequent consideration.
Therefore, by relabeling, we have $\Lambda^{1'}_8=\Lambda^1_8=\{R29, \textellipsis, R32\}$,
$\Lambda^{1'}_{10}=\Lambda^1_{10}=\{R33, \textellipsis, R36\}$,
$\Lambda^{2'}_{15}=\Lambda^2_{15}=\{R27, R28, R31, R32\}$ and
$\Lambda^{2'}_{16}=\Lambda^2_{16}=\{R34, R36, R38, R39\}$. The execution of
Step 3 produces $\Lambda^3_{20}=\{R26, R28, R30, R32\}$ \hspace {2mm} and
$\Lambda^3_{22}=\{R33, R34, R38, R40\}$.

{\bf Step 4}: Among the 24 rays in $\Lambda^{1'}_i$, $\Lambda^{2'}_i$
and $\Lambda^3_i$, two of them repeat 3 times and six of them
occur once. It means that there are only 8 different rays in
$\Lambda^{1'}_i$, $\Lambda^{2'}_i$ and $\Lambda^3_i$. These 8 rays constitute
the last base of the KS set generated by this algorithm. Adding the last base
will give us $\Sigma_{13}$ and $\Sigma_{14}$.

In our example, the rays that repeat 3 times in
$\Lambda^{1'}_8$, $\Lambda^{1'}_{10}$, $\Lambda^{2'}_{15}$,
$\Lambda^{2'}_{16}$, $\Lambda^3_{20}$ and $\Lambda^3_{22}$ are $R32$
and $R34$, while the rays that occur once are $R26$, $R27$, $R29$,
$R35$, $R39$ and $R40$. Therefore, the last base generated from Step
4 is $HB_{25}$, and we have $\Sigma_{13}=R32$ and $\Sigma_{14}=R34$.

The 13 bases generated by Algorithm II in our example are $PB_1$, $HB_6$,
$HB_7$, $HB_8$, $HB_{10}$, $PB_{2}$, $HB_{14}$, $HB_{15}$, $HB_{16}$,
$PB_{3}$, $HB_{20}$, $HB_{22}$ and $HB_{25}$ (following the order in which
they are generated). The 14 rays that occur 4 times are $R1$, $R2$,
$R3$, $R5$, $R9$, $R13$, $R14$, $R15$, $R19$, $R21$, $R23$, $R24$,
$R32$ and $R34$ (following the order in which they are chosen). The
remaining 24 rays occur twice in the generated 13 bases.

It can be seen from Algorithm II that to generate the KS sets of the
form $24_214_4\textendash 13_8$, 3 $\Gamma^{ij}$ that
determine $\Sigma_k (1 \leq k \leq 12)$ are required. All of the $\Gamma^{ij}$
in Table \ref{T2} are constructed from $PB$. There are 10 ways of choosing
3 out of 5 $PB$s, i.e., 123, 124, 125, 134, 135, 145, 234, 235,
245 and 345. Steps 1, 2 and 3 in Algorithm II show that there are 8, 4 and 2
ways of choosing first, second and third $\Gamma^{ij}$, respectively. Hence,
the total number of the KS sets in the type of $24_214_4\textendash13_8$, $N_{II}$,
that can be generated by Algorithm II is $ 10\times8\times4\times2=640$.

Finally, we explicates the steps taken in Algorithm III to generate the KS sets of the type
$20_220_4\textendash15_8$. The 15 bases are composed of 5 $PB$s and
10 $HB$s. The existence of 5 $PB$s means that 5 $\Gamma^{ij}$ are required.
Algorithm III shows us how to pick the 20 rays that
occur 4 times, $\Sigma_k$ ($k=1, \textellipsis, 20$), in such a way that
the remaining 20 rays occur twice in the KS sets generated.

{\bf Steps 1 to 3}: They are the same as Step 1 to  Step 3 of Algorithm II.

As an example, we take $\Gamma^{11}$, $\Gamma^{24}$ and
$\Gamma^{37}$ for the $\Sigma_k (1 \leq k \leq 12)$ as what happened in the first
3 steps of example for Algorithm II.  Again, at the end of the execution of
Step 3 we obtain $\Lambda^{1'}_8=\{R29, \textellipsis, R32\}$,
$\Lambda^{1'}_{10}=\{R33, \textellipsis, R36\}$, $\Lambda^{2'}_{15}=\{R27, R28, R31, R32\}$,
$\Lambda^{2'}_{16}=\{R34, R36, R38, R39\}$, $\Lambda^3_{20}=\{R26, R28, R30, R32\}$
and $\Lambda^3_{22}=\{R33, R34, R38, R40\}$.

{\bf Step 4}: Extract the fourth $\Gamma^{ij}$ from
$(\Lambda^{1'}_i)^4$, $(\Lambda^{2'}_i)^4$ and
$(\Lambda^3_i)^4$ to obtain $\Sigma_{13}, \textellipsis, \Sigma_{16}$.

Each of the $i$ in $\Lambda^{1'}_i$, $\Lambda^{2'}_i$ and
${\Lambda^3_i}$ takes two values. The symbols $(\Lambda^{1'}_i)^4$ refers to the
$\Lambda^{1'}_i$ which is a subset of $PB_4$. Similarly for
$(\Lambda^{2'}_i)^4$ and $(\Lambda^3_i)^4$. We obtain the fourth $\Gamma^{ij}$
from $[(\Lambda^{1'}_i)^4 \cap (\Lambda^{2'}_i)^4] \cup
[(\Lambda^{1'}_i)^4 \cap (\Lambda^3_i)^4] \cup
[(\Lambda^{2'}_i)^4 \cap (\Lambda^3_i)^4]$.

The execution of Step 4 generates the fourth $PB$ and another $HB$.

In our example, $\Gamma^{48} = [\Lambda^{1'}_8 \cap
\Lambda^{2'}_{15}] \cup [\Lambda^{1'}_8 \cap \Lambda^3_{20}] \cup
[\Lambda^{2'}_{15} \cap \Lambda^3_{20}]$. Thus, we have $\Sigma_{13} = R28$,
$\Sigma_{14} = R30$, $\Sigma_{15} = R31$ and $\Sigma_{16} = R32$. There are
only two newly generated bases, i.e., $PB_4$ and $HB_{24}$, because $HB_8$, $HB_{15}$
and $HB_{20}$ had been generated in the previous steps.

{\bf Step 5}: Extract the fifth $\Gamma^{ij}$ from
$(\Lambda^{1'}_i)^5$, $(\Lambda^{2'}_i)^5$ and
$(\Lambda^3_i)^5$ to obtain $\Sigma_{17}, \textellipsis, \Sigma_{20}$.

There is only one value remained for $i$ in $\Lambda^{1'}_i$, $\Lambda^{2'}_i$ and
${\Lambda^3_i}$ after the execution of Step 4. The symbols $(\Lambda^{1'}_i)^5$ refers to
$\Lambda^{1'}_i$ which is a subset of $PB_5$. Similarly for
$(\Lambda^{2'}_i)^5$ and $(\Lambda^3_i)^5$. We obtain the fifth $\Gamma^{ij}$
from $[(\Lambda^{1'}_i)^5 \cap (\Lambda^{2'}_i)^5] \cup
[(\Lambda^{1'}_i)^5 \cap (\Lambda^3_i)^5] \cup
[(\Lambda^{2'}_i)^5 \cap (\Lambda^3_i)^5]$.

The execution of Step 5 generates the fifth $PB$ and none of $HB$.

In our example, $\Gamma^{52} = [\Lambda^{1'}_{10} \cap
\Lambda^{2'}_{16}] \cup [\Lambda^{1'}_{10} \cap \Lambda^3_{22}] \cup
[\Lambda^{2'}_{16} \cap \Lambda^3_{22}]$. Thus, we have $\Sigma_{17} = R33$,
$\Sigma_{18} = R34$, $\Sigma_{19} = R36$ and $\Sigma_{20} = R38$. The only
generated base is $PB_5$ because $HB_{10}$, $HB_{16}$, $HB_{22}$ and $HB_{24}$ had
been generated in the previous steps.

The 15 bases generated by Algorithm III in our example are $PB_1$, $HB_6$,
$HB_7$, $HB_8$, $HB_{10}$, $PB_{2}$, $HB_{14}$, $HB_{15}$, $HB_{16}$,
$PB_{3}$, $HB_{20}$, $HB_{22}$, $PB_{4}$, $HB_{24}$ and $PB_{5}$
(following the order in which they are generated). The 20 rays that occur 4
times are $R1$, $R2$, $R3$, $R5$, $R9$, $R13$, $R14$, $R15$, $R19$,
$R21$, $R23$, $R24$, $R28$, $R30$, $R31$, $R32$, $R33$, $R34$,
$R36$, $R38$  (following the order in which they are chosen). The remaining
20 rays occur twice in the generated 15 bases.

Algorithm III requires 5 $\Gamma^{ij}$ to  generate the KS sets of
the form $20_220_4\textendash 15_8$. The 5 $\Gamma^{ij}$ chosen give
all the $\Sigma_k (1 \leq k \leq 20)$. The numbers of the ways of choosing the first,
second and third $\Gamma^{ij}$ in Step 1, Step 2 and Step 3 are 8, 4
and 2, respectively. On the other hand, there is only one way of choosing the fourth and
fifth $\Gamma^{ij}$ after the third $\Gamma^{ij}$ is chosen in Step
3. Thus the total number of KS sets in the type of
$20_220_4\textendash15_8$, $N_{III}$, that can be generated using
Algorithm III is $ 8\times4\times2=64$.

In summary, we have put forward three simple algorithms to completely
generate all of the KS sets in eight dimensional system with 36, 38 and 40 rays,
respectiely. Each of the algorithm is explained with the aid of an
example, and the total numbers of KS sets that can be generated using
these algorithms are 320, 640 and 64, respectively, which are agree
with the numbers reported in \cite{R18,R20}. Our algorithms also
provide the reasons for the existence of these numbers.

As our algorithms do not involve any computer calculation, they are
very helpful for those who lack of programming skill to
generate all of the 1024 KS sets or for those who need only few KS sets anytime and anywhere.
The steps in the algorithms are determined by how the rays that occur 4 times in the KS sets being chosen.
Table \ref{T2} lists the rays occurring 4 times that can be picked from pure bases ($PB$), while
some others are required to be chosen from hybrid bases ($HB$). To execute the algorithms, only
four types of sets of rays, denoted as $\Gamma^{ij}$, $\Lambda_i$, $\Delta_i$ and $\Xi^i_j$ need
to be extracted by inspection. Step 1 is common in all the three algorithms, whereas Step 2 and
Step 3 in Algorithm III are the same as for Algorithm II.
It is thus clear that the simplicity of the algorithms to generate
four-dimensional KS sets shown in \cite{R23} also exists in the case of eight-dimensional KS sets.

\begin{acknowledgments}
The author thanks B.A. Tay for valuable comments to improve this paper. This
work is supported by the Ministry of Higher Education of
Malaysia under the FRGS grant FRGS/1/2011/ST/UNIM/03/1.
\end{acknowledgments}


%




\begin{thebibliography}{99}
\bibitem{R1} Kochen S and Specker E P 1967 \emph{J. Math. Mech.} \textbf{17} 59
\bibitem{R2} Peres A 1991 \emph{J. Phys. A} \textbf{24} L175
\bibitem{R3} Conway J H and Kochen S, reported by Peres A 1993 \emph{Quantum Theory: Concepts and Method} (Dordrecht: Kluwer)
\bibitem{R4} Cabello A, Estebaranz J M and Garc\'{i}a-Alcaine G 1996 \emph{Phys. Lett. A} \textbf{212} 183
\bibitem{R5} Cabello A, Estebaranz J M and Garc\'{i}a-Alcaine G 2005 \emph{Phys. Lett. A} \textbf{339} 425
\bibitem{R6} Kernaghan M and Peres A 1995 \emph{Phys. Lett. A} \textbf{198} 1
\bibitem{R7} Meyer D A 1999 \emph{Phys. Rev. Lett.} \textbf{83} 3751
\bibitem{R8} Kent A 1999 \emph{Phys. Rev. Lett.} \textbf{83} 3755
\bibitem{R9} Cabello A 2008 \emph{Phys. Rev. Lett.} \textbf{101} 210401
\bibitem{R10} Kirchmair G, Z$\ddot{\mbox{a}}$hringer F, Gerritsma R, Kleinmann M, G$\ddot{\mbox{u}}$hne O, Cabello A,
Blatt R and Roos C F 2009 \emph{Nature} \textbf{460} 494
\bibitem{R11} Amselem E, Radmark M, Bourennane M and Cabello A 2009 \emph{Phys. Rev. Lett.} \textbf{103} 160405
\bibitem{R12} Moussa O, Ryan C A, Cory D G and Laflamme R 2010 \emph{Phys. Rev. Lett.} \textbf{104} 160501
\bibitem{R13} Badziag P, Bengtsson I, Cabello A and Pitowsky I 2009 \emph{Phys. Rev. Lett.} \textbf{103} 050401
\bibitem{R14} Aolita L, Gallego R, Ac\'{i}n A, Chiuri A, Vallone G, Mataloni P and Cabello A 2012 \emph{Phys. Rev. A} \textbf{85} 032107
\bibitem{R15} Cabello A 2010 \emph{Phys. Rev. Lett.} \textbf{104} 220401
\bibitem{R16} Horodecki K, Horodecki M, Horodecki P, Horodecki R, Pawlowski M and Bourennane M \emph{arXiv:quant-ph/1006.0468v1}
\bibitem{R17} Cabello A, D'Ambrosio V, Nagali E, Sciarrino F 2011 \emph{Phys. Rev. A} \textbf{84} 030302(R)
\bibitem{R18} Planat M 2012 \emph{EPJ Plus} \textbf{127} 86
\bibitem{R19} Harvey C, Chryssanthacopoulos J 2008 Worcester Polytechnic Institute, project number: PH-PKA-JC08
\bibitem{R20} Waegell M and Aravind P K 2012 \emph{J. Phys. A: Math. Theor.} \textbf{45} 405301
\bibitem{R21} Matsuno S 2007 \emph{J.\ Phys. A: Math. Theor.} \textbf{40} 9507
\bibitem{R22} Waegell M and Aravind P K 2013 \emph{Phys. Lett. A} \textbf{377} 546
\bibitem{R23} Waegell M and Aravind P K 2011 \emph{Found. Phys.}  \textbf{41} 1786
\end{thebibliography}
\end{document}